# Radiolysis of Amino Acids by Heavy and Energetic Cosmic Ray Analogs in Simulated Space Environments: α-Glycine Zwitterion Form

Williamary Portugal[1], Sergio Pilling[1*], Philippe Boduch[2], Hermann Rothard[2], Diana P. P. Andrade[1,2]

[1] – Universidade do Vale do Paraíba (UNIVAP), São José dos Campos, SP, 12244-000, Brazil.

[2] – Centre de Recherche sur les Ions, les Matériaux et la Photonique CIMAP (GANIL/CEA/CNRS/ENSICAEN/Université de Caen Basse-Normandie), F-14070 Caen Cedex 05, France

*e-mail: sergiopilling@pq.cnpq.br;

**ABSTRACT**

In this work, we studied the stability of the glycine molecule in the crystalline zwitterion form, known as α-glycine ($^{+}NH_3CH_2COO^{-}$) under action of heavy cosmic ray analogs. The experiments were conducted in a high vacuum chamber at heavy-ions accelerator GANIL, in Caen, France. The samples were bombarded at two temperatures (14 K and 300 K) by $^{58}Ni^{11+}$ ions of 46 MeV until the final fluence of $10^{13}$ ions cm$^{-2}$. The chemical evolution of the sample was evaluated in-situ using Fourrier Transformed Infrared (FTIR) spectrometer. The bombardment at 14 K produced several daughter species such as $OCN^{-}$, $CO$, $CO_2$, and $CN^{-}$. The results also suggest the appearing of peptide bonds during irradiation but this must be confirmed by further experiments. The half-lives of glycine in Interstellar Medium were estimated to be $7.8 \times 10^3$ years (300 K) and $2.8 \times 10^3$ years (14 K). In the Solar System the values were $8.4 \times 10^2$ years (300 K) and $3.6 \times 10^3$ years (14 K). It is believed that glycine could be present in space environments that suffered aqueous changes such as the interior of comets, meteorites and planetesimals. This molecule is present in proteins of all alive beings. So, studying its stability in these environments provides further understanding about the role of this specie in the pre-biotic chemistry on Earth.

## 1 INTRODUCTION

The interstellar medium (ISM), a vast space among stars, is a rich reservoir of the gas and dust as well as organic compounds (e.g. Ehrenfreund & Charnley 2000). Denser regions of ISM, called molecular clouds, are characterized by very low temperatures in the order 10-30K with density around $10^4$-$10^8$ particles cm$^{-3}$. Due to the low temperature, molecules from gas-phase are adsorbed on the surface of dust grains (aggregates of non-voltatile species such as carbonaceous species, oxides and silicates) producing a water-rich ice mantle. In addition to amorphous water, interstellar ices consist of a variety of simple molecules such as $CO_2$, $CO$, $CH_3OH$ and $NH_3$ (e.g. Boogert & Ehrenfreund, 2004).

Laboratory studies have shown that the photolysis and radiolysis of space ice analogs can produce complex organic compounds and pre-biotic molecules such as amino acids and nucleobases (e.g. Bernstein et al., 2002; Muñoz Caro et al., 2002; Kobayashi et al., 2008; Pilling et al., 2009). The presence of these pre-biotic compounds was also observed in laboratory investigations of meteorite samples, where more of the 70 amino acids were identified, as in the case of the Murchison meteorite (Cronin & Pizzarello 1983; Cronin, Pizzarello & Cruikshank, 1988; Glavin & Dworkin, 2009; Glavin et al., 2011).

Glycine ($NH_2CH_2COOH$), the simplest amino acid present in several proteins and enzymes in all forms of life on Earth, is predicted to exist in space (e.g. Kuan et al., 2003). Although this specie has not been observed yet in ISM, it was detected together with some of its precursors, by chromatographic analysis and carbon isotopes technique, in the samples collected from the 81P/Wild 2 comet and delivered to Earth by NASA Stardust spacecraft (Elsila, Glavin & Dworkin, 2009). This study reinforces the ideas of the other authors (e.g. Chyba et al. 1990; Cronin, Pizzarello & Cruikshank, 1998) that great amounts of organic materials could have been brought to Earth by comet and asteroid impacts. The research about this specie helps the understanding about pre-biotic chemistry, which could have been responsible for the emergence of the life on early Earth. Glycine in aqueous form is mainly zwiterionic glycine ($^+NH_3CH_2COO^-$) (Pilling et al., 2013). However, in the condensed phase it is crystallized in different structures. According Liu et al. (2008), it can be found in three crystalline forms (α, β and γ-glycine), which are different from each other due to the angle between C-N bond and vibration mode around C-C bond.

According to those authors, these three glycine forms are the consequence of the arrangement of the molecule on crystal structure and several kinds of interactions: a) Van der Waals, b) electrostatic and c) hydrogen atoms bond. The hydrogen bonds play an essential role in the organization of the crystallized glycine structure. According Pilling et al. (2013), in space, the glycine molecule can be formed by reactions gas phase or in the ice surface by induced reactions. Radio observations have suggested that the abundance of glycine in space (gas phase) must be very low. But, if we consider that small amount of glycine is present in ISM and if it is continuously being produced from gas phase and deposited onto grains or produced directly on grain surfaces (e.g. Woon, 2002; Zhu & Ho, 2004; Pilling et al., 2011a), its abundance on grains will increase with time during the evolution of an interstellar or protostellar cloud. Futhermore, the glycine desorbs only above 350 K in high vacuum conditions (e.g. Pilling et al., 2011). Then, if such grains were exposed at temperatures around 200 K, all volatile compounds including water will be desorbed and the amount of glycine will increase with time, leading to the production of localized crystalline structures, such as β-glycine or α-glycine. These crystalline forms may exist in space depending of the of hydration degree as discussed by Pilling et al. (2013). For example, due to radioactivity inside comets a fraction of water may exist in liquid phase, allowing the presence of α-glycine. The β-glycine form can be found in warm environments, HII regions and on the dusts of Difuse Intestellar Medium (DISM), where the presence of water is absent. In astrophysical environments, at different temperatures, the glycine is exposed to several ionizing radiation fields (eg. UV, X-rays and cosmic rays).

As discussed at Pilling et al., 2011a, the formation of amino acids from simple compounds such as CO, $CO_2$, $CH_4$, $NH_3$, $H_2O$ and $H_2$ (and others) have been performed for at least 60 years in an attempt to simulate primitive Earth conditions (e.g., Miller, 1953, 1955; Sanchez, Ferris, & Orgel, 1966; Ponnamperuma & Woeller, 1967; Zhu & Ho, 2004) or interstellar/protoplanetary grain mantles (e.g., Bernstein et al., 2002; Munõz Caro et al., 2002; Holtom et al., 2005; Elsila et al., 2007; Nuevo et al., 2008; Pilling et al., 2010a; de Marcellus et al., 2011). In most of those experiments, the gas mixtures or a film produced by the frozen gas mixtures are submitted to ionizing radiation sources (electrons, photons, fast ions and swift heavy ions), which trigger the physical-chemistry reactions that allows complex molecules to be formed. The formation of glycine was also the subject of several theoretical studies involving different reactions set in the gas phase (e.g., Blagojevic, Petrie & Bohme, 2003; Largo, Redondo & Barrientos, 2003; Maeda & Ohno, 2004; Bossa et al. 2009; Largo et al., 2010) and in or on interstellar grain analogues (e.g., Sorrell, 2001; Woon, 2002; Mendoza et al., 2004; Rimola & Ugliengo, 2009).

In this work, we simulated in laboratory the physical-chemical conditions found in space environments such as in the surface of grains inside denser regions of ISM as well as grains in protostellar disks, comets and asteroids. The current experiments investigate the stability of glycine molecules in condensed phase (crystalline zwiterionic form α-glycine) against heavy ion cosmic ray analogs. The cross sections obtained empirically here are compared with values obtained recently by Pilling et al. (2013) and Gerakines et al. (2012) employing light ions (~1 MeV protons) at different temperatures.

In section 2, we present details about the experimental methodology. Results are presented in section 3. The results are discussed in section 4, including the determination of the dissociation cross section and formation of daughter species (sample at 14 K). Astrophysical implications, involving the half-lives determination in different space environments with radiation ionizing field are discussed in section 5. A summary containing primary conclusions is in section 6.

## 2 EXPERIMENTAL METHODOLOGY

The experiments were perfomed in a stainless steel chamber, under high vacuum conditions, coupled to the experimental beam line IRRSUD (IRRadiation SUD) of the heavy-ion accelerator GANIL (Grand Accélérateur National d'Ions Lourds), in Caen, France. The α-Glycine crystal samples were prepared outside the vacuum chamber by drop casting of glycine aqueous solution (0.1 M) on CsI substrate. The sample was commercially bought from Sigma-Aldrich and had a purity of 99.9%. The experiments were done at two different temperatures: 14 K and 300 K. The first sample was introduced inside the vacuum chamber and was bombarded at 300 K. The second sample was inserted into the chamber at 300 K and cooled to 14 K. After the bombardment at 14 K, the sample was slowly heated to the room temperature. During this process, several infrared spectra were obtained from the sample to evaluate the changes due the heating. During the experiment, the chamber pressure was around $2 \times 10^{-8}$ mbar.

The samples were irradiated by $^{58}$Ni$^{11+}$ ions at energy of 46 MeV. The flux was $2 \times 10^{9}$ ions cm$^{-2}$ s$^{-1}$. The incoming charge of $q = 11^{+}$ is close to the mean charge of the ions inside the solid sample after several encounters with atoms in the target (Pilling et al., 2010a). The infrared spectra of glycine before and after different irradiation fluencies (ions cm$^{-2}$) were obtained *in-situ* with Fourier Transform Infrared (FTIR) spectrometer (Nicolet - Magna 550), with wavenumbers ranging from 4000 to 650 cm$^{-1}$ and with 1 cm$^{-1}$ resolution. The spectrum of the clean CsI substrate was measured for background subtraction. Figure 1 presents a schematic diagram of the experiment.

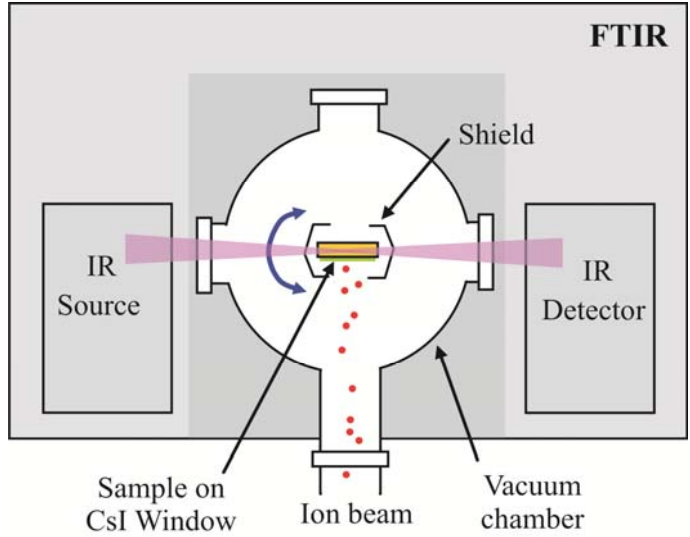

**Figure 1.** Schematic experimental setup employed for bombardment of α-glycine crystal by heavy ions.

The molecular column density $N$ was obtained from the relation between optical depth $\tau_v = \ln(I/I_0)$ and the band strength $A$ (cm molec$^{-1}$), of the respective vibration mode of a given species. Here, $I_0$ and $I$ are the intensity of light before and after passing through a sample, respectively. Since the absorbance measured by FTIR is given by $Abs_v = \log(I_0/I)$, the column density of the samples according to Pilling et al. (2010a) is calculated by

$$N = \frac{1}{A}\int \tau_v dv = \frac{2.3}{A}\int Abs_v dv \qquad [molecules\ cm^{-2}] \qquad [1]$$

where $Abs_v = \ln(I/I_0)/\ln(10) = \tau_v/2.3$.

As discussed by Pilling et al. (2011) the thickness of the sample can be estimated by the expression

$$d = \frac{N_0}{6.02\ x\ 10^{23}}\frac{M}{\rho}\ x\ 10^4 \qquad [\mu m] \qquad [2]$$

where $N_0$ is the initial column density in molecules cm$^{-2}$, $M$ is the molar mass in g mol$^{-1}$ and $\rho$ is the glycine density in g cm$^{-3}$. In this work we adopted the absorbance of the CN stretch vibration mode of glycine to determine the column density of the sample. The obtained value for the initial column density of α-glycine at 14 K and 300 K, was 1 x 10$^{18}$ cm$^{-2}$ and 0.3 x 10$^{18}$ cm$^{-2}$, respectively. Assuming 1.16 g/cm$^3$ as the glycine average density in both samples, the sample thickness was estimated to be around 1-2 μm.

Due to the high velocity of the incident $^{58}$Ni$^{11+}$ ions, the energy deposition on the glycine sample is mainly by inelastic interactions with target electrons (electronic stopping power) (Pilling et al., 2010a). Despite the ion flux employed in laboratory experiments to be several orders of magnitude higher than the flux of similar ions in space and the thickness of laboratory ice also is being higher, the energy delivered and the damage induced by similar ions in both scenarios have great similarity. Considering the energy of 46 MeV, we calculate, by means of SRIM - Stopping and Ranges of Ions in Matter software, the electronic and nuclear energy loss of the projectiles in the glycine sample, which were 5270 keV μm$^{-1}$ and 16,1 keV μm$^{-1}$, respectively. The maximum penetration depth of ions in the glycine sample was calculated to be 17.5 μm. In this case, the electronic stopping power is at least two orders of magnitude greater than the nuclear one. Then, the depth of penetration being one order of magnitude greater than the thickness of the sample, the ions cross the entire sample losing only about 10% of the their initial energy ensuring an inelastic collision regime. These assumptions are in good agreement with cosmic ray impact in submicron ices in interstellar or interplanetary regions as discussed by Pilling et al. (2012).

In a similar experiment performed by Pilling et al. (2010a), the authors discussed that each projectile induces significant changes in the sample only in a region with roughly 3 nm of diameter. According Pilling et al. (2012), considering a constant and homogeneous ion flux of $2 \times 10^9$ ions cm$^{-2}$ s$^{-1}$, the average distance between two nearby impacts is roughly 300 nm (about 100 times higher than the length of processed sample by a single projectile hit). From these values, the authors estimate that the probability of an ion to hit a given area of 3 nm in diameter each second is about $P \sim 1.4 \times 10^{-4}$. Therefore, only after about 2 h of continuous bombardment a second projectile hit in the same nanometric region. This is enough time to consider that each projectile hit always a thermalized region in the laboratory. According Pilling et al. (2012), such scenario is very similar to interstellar or interplanetary regions which have very low ion flux as well as ion fluence (even integrated over a large extended period of time). In this work, the maximum fluence (1 x 10$^{13}$ ions cm$^{-2}$) employed on the bombardment of the sample was obtained after 1.5 hour of direct exposure to cosmic ray analogs. Such fluence corresponds to the 10$^6$ years of bombardment in the ISM considering the flux of heavy ions (12 ≤ Z ≤ 29) with energy between 0.1 to 10 MeV in the interstellar medium estimated by Pilling et al. (2010a; 2010b) being about 5 x 10$^{-2}$ cm$^{-2}$ s$^{-1}$.

## 3 RESULTS

The spectra obtained from non-irradiated and irradiated α-glycine samples at different fluencies by $^{58}$Ni$^{11+}$ ions are shown in the Figures 2a and 2b, respectively for the sample at 14 K and 300 K. Using these spectra, we can compare the effect of the different fluencies on the sample. In both figures, the non-irradiated crystalline sample is represented in the top spectrum. The arrow on the peak at 1034 cm$^{-1}$ indicates the stretching vibrational mode of the C-N bond on the glycine molecule, employed to quantify the sample by means of equation 1. Beyond structural differences, crystals of α e β- glycine exhibit differences to in their infrared spectrums for example, band displacements (see details in Pilling et al. 2013). In the stretching vibration mode of the C-N, α-glycine has a center band at 1034 cm$^{-1}$ and 1040 cm$^{-1}$ for β-glycine.

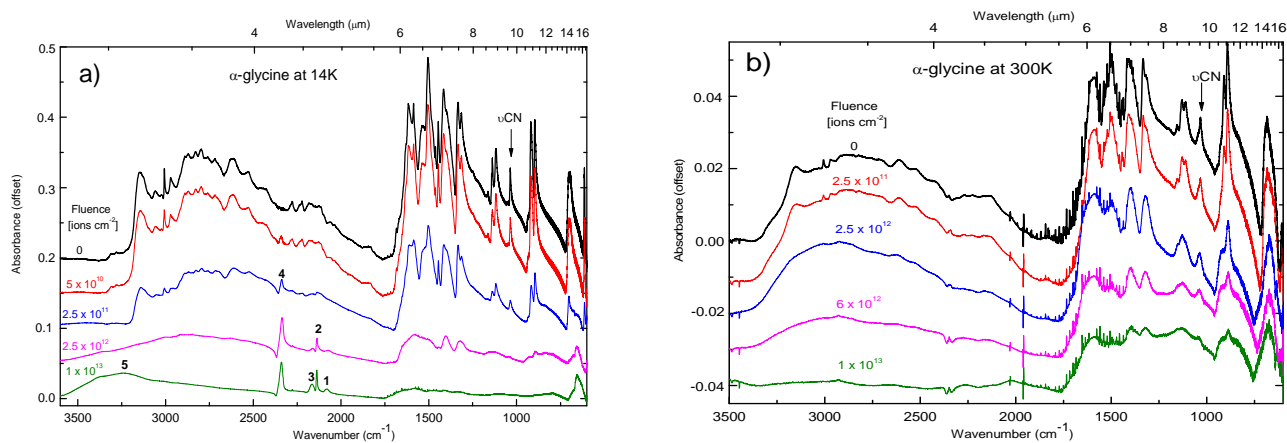

**Figure 2.** Infrared spectra of α-glycine ($^+NH_3CH_2COO^-$) before (top dark line) and after different irradiation fluencies. The arrow on the peak at 1034 cm$^{-1}$ indicates the location of CN stretching mode employed to quantify the sample. The peak of daughter species formed on the irradiated sample at 14 K are indicated by numbers: 1 – CN$^-$ (2080 cm$^{-1}$); 2 – CO (2137 cm$^{-1}$); 3 - OCN$^-$ (2165 cm$^{-1}$); 4 - $CO_2$ (2336 cm$^{-1}$); 5 - $H_2O$ (3280 cm$^{-1}$). a) sample at 14 K b) sample at 300K.

The peak position and band strengths of some vibration modes of non-bombarded α-glycine crystal at 14 K is listed in Table 1 (taken from Holton et al. 2005). In the glycine sample at 14 K, the bands at 2080 cm$^{-1}$, 2137 cm$^{-1}$, 2165 cm$^{-1}$, 2336 cm$^{-1}$ and 3280 cm$^{-1}$ of Figure 2a represent the daugher species CN$^-$, CO, OCN$^-$, $CO_2$ and $H_2O$, respectively, formed from irradiation of the sample (Pilling et al., 2010a; Gerakines et al., 1995). These species were produced from chemical processes triggered by heavy ions induced radiolysis. In the experiment at 300 K, new species could also be formed during the bombardment, however, due to the relatively high temperature of the sample, volatile species such as CO, $CO_2$ and $H_2O$ may be desorbed and thus they were not found in the spectrum. The $H_2O$ observed among the daughter species may also have contribution of the adsorption of residual gas in the backside of the CsI substrate Since the partial pressure of this compound inside vacuum chamber is of the order of $1\times10^{-8}$ mbar (~ 1 monolayers adsorbed after each 100 seconds).

**Table 1.** Peak position (in cm$^{-1}$ and μm) and band strengths **(also called A-value)** of some vibration modes of the non-bombarded α-glycine crystal at 14 K (taken from Holton et al. 2005).

| Position cm$^{-1}$ | Position μm | Assigment | Characterization | A-Value (cm molecule$^{-1}$) |
|---|---|---|---|---|
| 3076, 3040 | 3.2, 3.3 | $\nu_{as}$ NH$_3$ | Asymmetric stretch | 2.17 x 10$^{-17}$ |
| 1596 | 6.3 | $\nu_{as}$ CO$_2^-$ | Asymmetric stretch | 8.77 x 10$^{-17}$ |
| 1505 | 6.6 | $\delta_s$ NH$_3$ | scissoring | 4.39 x 10$^{-17}$ |
| 1413 | 7.1 | $\nu_s$ CO$_2^-$ | Symmetric stretch | 3.86 x 10$^{-17}$ |
| 1334 | 7.5 | ω CH$_2$ | Wagging | 8.77 x 10$^{-17}$ |
| 1131 | 8.8 | NH$_3$ | stretching | 1.43 x 10$^{-18}$ |
| 1112 | 8.9 | ρNH$_3$ | rocking | 4.70 x 10$^{-18}$ |
| 1034 | 9.7 | ν CN | stretching | 1.38 x 10$^{-18}$ |

As described by Pilling et al. (2010a), the variation of the molecular abundance in the sample during the bombardment can be assigned to several process: i- molecular dissociation (quantified by dissociation cross-section); ii- molecular desorption (molecules sublimated from the solid sample due to incident radiation); iii- formation of the new molecules (quantified by formation cross-section).

4 DISCUSSION

4.1 DISSOCIATION CROSS-SECTION OF GLYCINE UNDER BOMBARDMENT OF NI IONS

The dissociation (or destruction) cross section ($\sigma_d$) of a given molecule under a given dissociation processes is unique and should be reproducible at wherever laboratory. However, when we use the evolution of bands in the infrared spectra (molecular vibration modes) to determine such value we may have different approaches for that, as we will discuss further. In this work, we present another kind of cross section called specific bond rupture cross section, that is conceptually different than the molecular dissociation cross section and monitors the sensitivity of a given molecular bond to a give dissociation processes such as the ion bombardment.

The specific bond rupture cross-sections $\sigma_{(XY)}$ of given molecule can be obtained from the expression

$$\ln\left(\frac{a}{a_0}\right) = -\sigma_{(XY)} \cdot F \qquad [3]$$

where $a_0$ and $a$ are the integrated absorbance of a given vibration mode involving X and Y atoms observed in the infrared spectrum at the beginning of the experiments and at a given fluence, respectively. $F$ is the fluence in units of ions cm$^{-2}$.

The degradation of the sample under action of $^{58}$Ni$^{11+}$ ions of 46 MeV/u is presented quantitatively in the figures 3a and 3b for the experiments at 14 and 300 K, respectively. These figures show the peak area dependence as a function of the fluence (normalized to the initial peak area) for three different functional groups on the glycine molecule: CN, CH e NH. Since different bonds in the glycine molecule have different bond energies, they can have also different fragility under irradiation (or ionizing field) as illustrated by different dissociation cross section of specific bonds. This difference in the dissociation cross-section for specific molecular bonds in solid sample was also observed in similar experiments in the literature (e.g. Pilling et al. 2013; Andrade et al. 2013 **and Bergantini et al. 2014**).

Bellow we list there different methodologies that can be adopted for the determination of the molecular dissociation cross section ($\sigma_d$) of a given molecule directly from the evolution of bands in the infrared spectrum as function of the fluence ionizing radiation:

i) We can supposed that the molecular dissociation cross section is ruled by the maximum value obtained for the specific bond rupture cross section ($\sigma_{(XY)}$). This can be expressed as

$$\sigma_d = \max_{i=1,n} \sigma_{(XY)_i} \qquad [4]$$

However, in some cases, this may introduce an additional error if we considered atoms that have very low bond energy (such as hydrogen atoms in sp$^3$ orbital, or CH bond) that can be easily interchanged with the surrounding molecules (matrix/solvent/substrate).

ii) An alternative way is to suppose that the molecular dissociation cross section is ruled by the value obtained for the cross section of the rupture of a molecular bond within the molecular backbone, $\sigma_{(XY)\,backbone}$. This can be written as

$$\sigma_d = \sigma_{(XY)backbone} \qquad [5]$$

An example of molecular backbone bond in glycine molecule is the CN bond (this methodology has been employed by Pilling et al. 2013). The value of the dissociation cross section of glycine, considering the C-N bond rupture and the action of Ni ions, at low temperature is about 23 times higher than at room temperature. This may be due to the high density in the sample at 14 K as pointed out previously by Pilling et al. (2013). Another explanation for this enhancement in the dissociation at low temperature is the possible reaction among new formed species with each other and others already presented in the sample. The C-N bond in the α-glycine molecule at 14 K was the most sensible bond during the ion bombardment. Curiously, this bond is the less sensitive in the sample at 300 K. For comparison, the values obtained by Pilling et al. (2013) employing 1 MeV protons on α-glycine crystals at 300 K are also given in the Table 2. Comparing only the data at 300 K bombarded by different projectiles, we observe that the sensitivity of N-H bonds is much higher than that of other groups, when heavy ion is employed. For proton bombardment the molecular bonds are destroyed in a similar way in both cases.

iii) However, the simply way to derive the molecular dissociation cross section from measurements in the infrared is to consider the average value between the different specific bond rupture cross sections (or in some cases by the evolution of all the area in the infrared band). In this work we adopted this methodology analyzing (and we analyzed) the evolution of the rupture of three molecular bonds (CN, NH and CH) of the glycine molecule in the infrared spectrum during the bombardment with 46 Mev Ni ions. Therefore, we write the glycine molecular dissociation cross section as

$$\sigma_d = \sum_{i=1,n}^{n} \frac{\sigma_{(XY)_i}}{n} = \frac{\sigma_{(CN)} + \sigma_{(NH)} + \sigma_{(CH)}}{3} \qquad [6]$$

Considering this methodology, the obtained value at 14 K and 300 K were $\sigma_d = 2.4\times 10^{-12}$ cm$^2$ and $3.4\times 10^{-13}$ cm$^2$, respectively. Such methodology was also employed by Pilling et al., 2011c at experiments on the processing of solid glycine by ionizing photons. The error in the determination of specific bond rupture cross section lay in the uncertainty of the area determination in the infrared spectra that was below 10%. In this work, the estimated error for glycine dissociation cross section ($\sigma_d$), was adopted equal to the standard error from the determined specific bond rupture cross sections. Such value was about 20% and 50% for the experiments at 14 K and 300 K, respectively. The values of specific bond rupture cross section and the glycine dissociation cross section are listed in Table 2. The value of the dissociation cross section considering the average value at low temperature is about seven times higher than at room temperature. It is possible to see in the Table 2 that for any adopted vibration mode or considering the average values, the glycine dissociation cross sections are higher for the sample at 14 K.

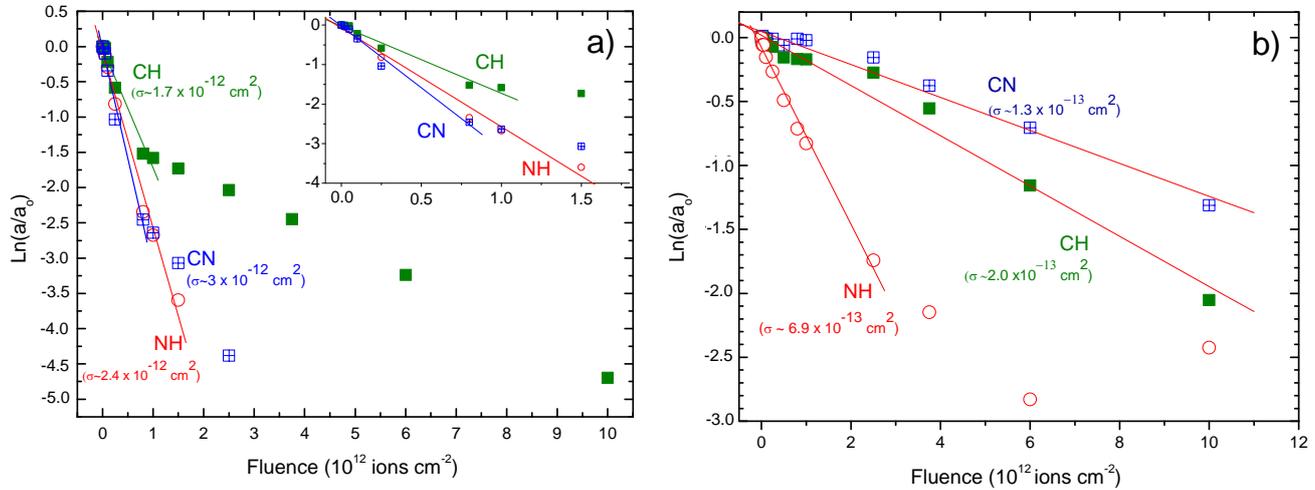

**Figure 3.** Evolution of the integrated absorbance of selected vibration modes (CN, CH, and NH) of α-glycine crystal as a function of ion fluence. a) sample at 14 K b) sample at 300 K. Lines represent the best fitting employing Eq. [3]. The determined values of the cross-sections for specific bond ruptures are indicated in the Figure.

Table 2. Cross-sections for specific bond ruptures $\sigma_{(CN)}$, $\sigma_{(C-H)}$ and $\sigma_{(N-H)}$ of glycine molecule and glycine dissociation cross section ($\sigma_d$) due to the bombardment of 46 MeV $Ni^{11+}$ at 14 K and 300 K. For comparison purpose, the value obtained by Pilling et al. (2013) employing 1 MeV protons at 300 K are also given.

| Cross Sections | Vibration mode | Value at 14 K ($10^{-12}$ cm$^2$) | Value at 300 K ($10^{-13}$ cm$^2$) | Ratio | Value obtained by Pilling et al. 2013. ($10^{-14}$ cm$^2$) |
|---|---|---|---|---|---|
| $\sigma_{(CN)}$ | C-N ($\nu$ CN) | ~3.0 ± 0.3 [a] | 1.3 ± 0.1 [a] | 23.1 | 2.5 |
| $\sigma_{(CH)}$ | C-H ($\omega$ CH$_2$) | 1.7 ± 0.2 [a] | ~2.0 ± 0.2 [a] | 8.5 | 2.1 |
| $\sigma_{(NH)}$ | N-H ($\nu_{as}$ NH$_3$) | 2.4 ± 0.2 [a] | 6.9 ± 0.7 [a] | 3.5 | 3.1 |
| $\sigma_d$ | - | 2.4 ± 0.4 [b] | 3.4 ± 1.8 [b] | 7.1 [c] | 2.6 [d] |

[a] The estimated errors are from the uncertainty in the band area determination (~10%). [b] The molecular dissociation cross section ($\sigma_d$) is given by the average value for the dissociation cross section of specific bond rupture, as discussed in the text. The presented errors are given by standard error of the values determined from cross section of specifics bonds ruptures;
[c] Obtained from the average values of the ratios listed above; [d] Average value for $\sigma_d$ calculated from Pilling et al. 2013.

## 4.2 MODELING THE DISSOCIATION CROSS SECTION OF GLYCINE UNDER ACTION OF ALL ENERGETIC IONS

In this work, the half-lives of glycine molecule were estimated considering the interaction with the several cosmic ray constituents. **As discussed previously by Andrade et al. (2013), de Barros et al. (2014) and references therein, the dissociation cross section of a frozen molecule under bombardment by heavy ions can be modeled by the power law**

$$\sigma_d = a\, S_e^{\,n} \qquad [7]$$

where $\sigma_d$ **is the molecular dissociation cross-section in $cm^2$, $S_e$ is the electronic stopping power and $a$ and $n$ are constants empirically determined.** According to those authors, when the ionic projectiles pass through the target material, they continually transfer their energies to target and also induce ionization process, resulting in second generation of ions, radicals, electrons, photons and excited species. For ion beams with high velocities, the projectile energy and momentum are mainly transferred to the target by ion electron interaction (electronic stopping regime). The projectiles traverse each molecular layer, losing a small amount of their energy along the pathway, modifying its charge state and velocity. The electronic stopping power ($S_e$) was calculated using the SRIM code written by Ziegler et al. **(2008).**

**Figure 4 shows the dependence of the glycine dissociation cross-section on the electronic stopping power at two temperatures (14 and 300 K). The black circles represent the average values for the cross section obtained employing $^{58}Ni^{11+}$ ion, at 14 and 300 K (this work). The red bars illustrate the range of the values determined for the specific bonds dissociation cross sections in this work.** For H ions we used the $S_e$ value for 0.8 MeV and 1 MeV at 14 K and 300 K, respectively, and the cross-section of $2.9 \times 10^{-15}$ $cm^2$ and $2.5 \times 10^{-14}$ $cm^2$, according to data of Pilling et al. (2013) and Gerakines et al. (2012). The cross section estimated for H ions ($2.9 \times 10^{-15}$ $cm^2$) from Gerakines et al. (2012) is for glycine plus $H_2O$ sample. In this work, we adopted that the dissociation cross-section of glycine by Ni ions were the average value of three specific bonds dissociation cross-section, as listed in Table 2. Therefore the glycine dissociation cross-section were $2.4 \pm 0.4 \times 10^{-12}$, $3.4 \pm 1.8 \times 10^{-13}$ **$cm^2$, at 14 K and 300 K respectively. Lines show the best fittings employing equation 7 for the two set of data at two distinct temperatures. The values of the constants obtained from the models are indicated in the figure and also listed in the Table 3.**

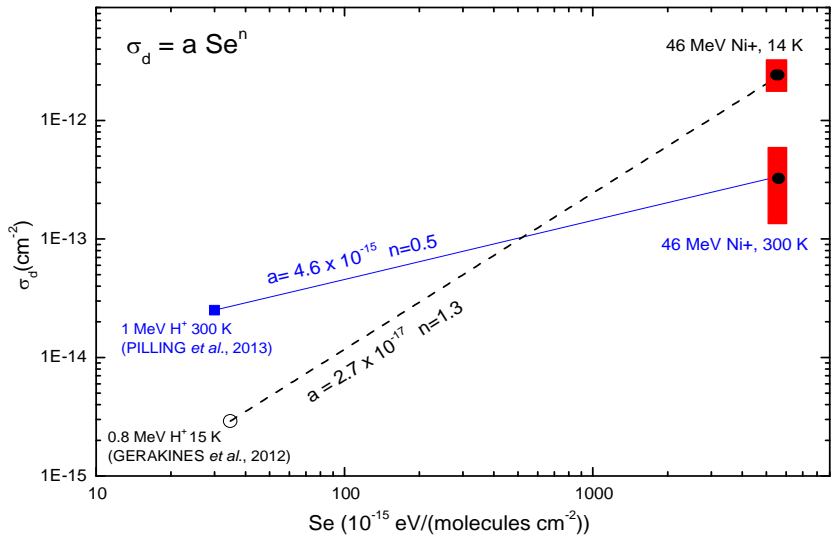

**Figure 4.** Dependence of glycine dissociation cross-section on the electronic stopping power at two temperatures (14 and 300 K). The black circles represent the average values for the cross section obtained from data of $^{58}Ni^{11+}$ ion, at 14 and 300 K in this work. The red bars illustrate the range of the values determined for the specific bonds dissociation cross sections in this work.

**Table 3.** Best fit parameters from the equation of the dissociation cross section as function of electronic stopping power: $\sigma_d = a \cdot Se^n$.

| Temperature (K) | n   | a                    |
|-----------------|-----|----------------------|
| 14              | 1.3 | $2.7 \times 10^{-17}$ |
| 300             | 0.5 | $4.6 \times 10^{-15}$ |

The *a* and *n* values of **listed in** Table 3 were used to calculate the destruction cross-sections ($\sigma_d$) of glycine molecules at 14 K and 300 K under bombardment by several ions according to equation [7]. The destruction cross sections as function of energy, at 14 K and 300 K respectively, are presented in Figs. 5a and 5b. These figures show that in the range between approximately 0.1-10 MeV/u, cosmic rays promote the higher destruction in the samples. In addition, in this energy range, the average of $\sigma_d$ is higher at 14 K than at 300 K. For very higher energies we observe the opposite behavior. This can to explain the low variation among the half-lives of the sample at 14 K, leading in consideration energy range of ~0.1-1.5x10$^3$ MeV/u and ~0.1-10 MeV. The initial energy studied for each ion is the energy which $S_e$ is at least ten times higher than $S_n$ (nuclear stopping power) on the same energy range to all ions. This energy range is ~0.1 MeV/u - 1.5 x 10$^3$ MeV/u. The highest destruction cross-section rate to the most ions (heavy ions) in the samples at both temperatures (14 and 300 K) occurs about 1 MeV/u. It is possible realize also, that the dissociation cross sections of glycine irradiated by H and He, as a function of energy, are lower at 14 K than at 300 K. The glycine dissociation cross section considering different ions (main constituents of cosmic rays inventory) in selected energies (0.1, 10 and 1000 MeV/u) as well the average value are presented in the Table 4.

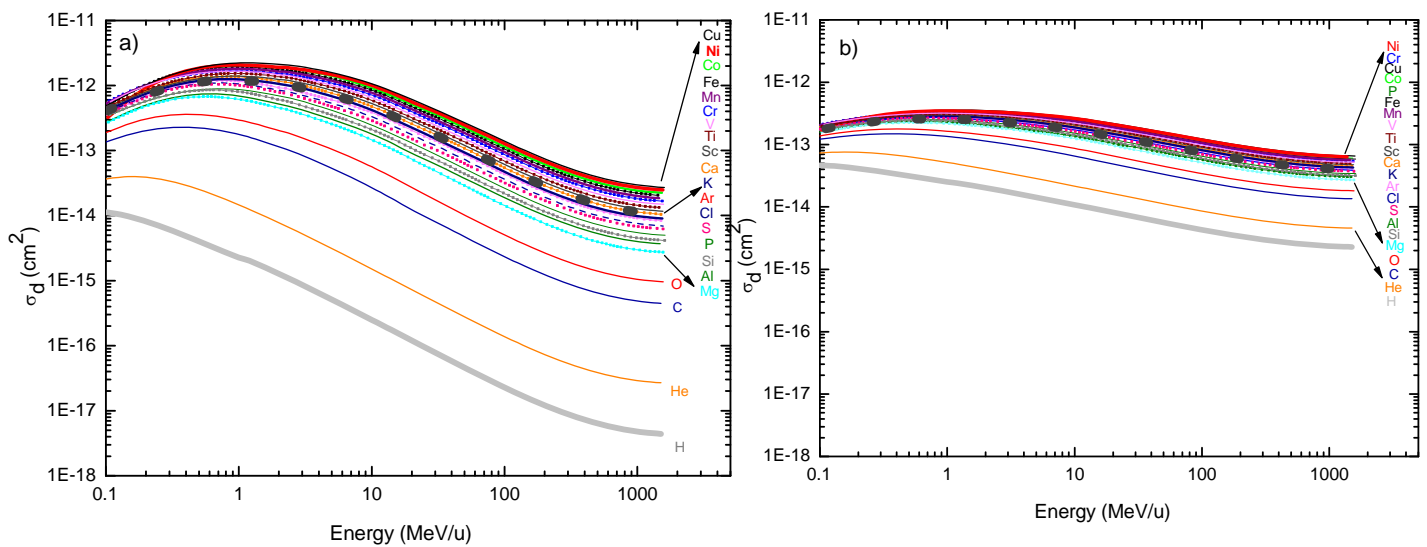

**Figure 5.** Destruction cross-section of glycine by cosmic rays as function of energy. a) For the experiment at 14 K and b) for 300 K. See details in text. The large dashed line, in both graphics, is the average value.

**Table 4.** Glycine dissociation cross section considering different ions (main constituents of cosmic rays inventory) in selected energies (0.1, 10 and 1000 MeV/u) as well the average value. The estimated error at 14 K was around 20% and at 300 K was around 50%. Such error values were mainly ruled by the standard error of the glycine dissociation cross-section employing Ni ions as illustrated in Table 2.

| | 0.1 MeV/u | | 10 MeV/u | | 1000 MeV/u | |
|---|---|---|---|---|---|---|
| Ions | 14K | 300K | 14K | 300K | 14K | 300K |
| Al | 2.8 x 10$^{-13}$ | 1.6 x 10$^{-13}$ | 1.7 x 10$^{-13}$ | 1.4 x 10$^{-13}$ | 4.0 x 10$^{-15}$ | 3.1 x 10$^{-14}$ |
| Ar | 3.6 x 10$^{-13}$ | 1.8 x 10$^{-13}$ | 3.8 x 10$^{-13}$ | 1.8 x 10$^{-13}$ | 9.1 x 10$^{-15}$ | 4.3 x 10$^{-14}$ |
| Ca | 4.6 x 10$^{-13}$ | 2.0 x 10$^{-13}$ | 4.8 x 10$^{-13}$ | 2.0 x 10$^{-13}$ | 1.1 x 10$^{-14}$ | 4.7 x 10$^{-14}$ |
| Cl | 3.8 x 10$^{-13}$ | 1.8 x 10$^{-13}$ | 3.3 x 10$^{-13}$ | 1.7 x 10$^{-13}$ | 7.5 x 10$^{-15}$ | 4.0 x 10$^{-14}$ |
| Co | 2.0 x 10$^{-12}$ | 3.4 x 10$^{-13}$ | 9.1 x 10$^{-13}$ | 2.5 x 10$^{-13}$ | 2.4 x 10$^{-14}$ | 6.3 x 10$^{-14}$ |
| Cr | 5.8 x 10$^{-13}$ | 2.1 x 10$^{-13}$ | 7.4 x 10$^{-13}$ | 2.3 x 10$^{-13}$ | 1.8 x 10$^{-14}$ | 5.6 x 10$^{-14}$ |
| Cu | 5.0 x 10$^{-13}$ | 2.0 x 10$^{-13}$ | 1.0 x 10$^{-12}$ | 2.7 x 10$^{-13}$ | 3.0 x 10$^{-14}$ | 6.8 x 10$^{-14}$ |
| Fe | 5.2 x 10$^{-13}$ | 2.0 x 10$^{-13}$ | 8.6 x 10$^{-13}$ | 2.5 x 10$^{-13}$ | 2.2 x 10$^{-14}$ | 6.1 x 10$^{-14}$ |
| K | 4.2 x 10$^{-13}$ | 1.9 x 10$^{-13}$ | 4.1 x 10$^{-13}$ | 1.9 x 10$^{-13}$ | 9.7 x 10$^{-15}$ | 4.4 x 10$^{-14}$ |
| Mg | 2.7 x 10$^{-13}$ | 1.6 x 10$^{-13}$ | 1.4 x 10$^{-13}$ | 1.2 x 10$^{-13}$ | 3.0 x 10$^{-15}$ | 2.8 x 10$^{-14}$ |
| Mn | 5.3 x 10$^{-13}$ | 2.0 x 10$^{-13}$ | 7.8 x 10$^{-13}$ | 2.4 x 10$^{-13}$ | 2.0 x 10$^{-14}$ | 5.8 x 10$^{-14}$ |
| Ni | 3.2 x 10$^{-13}$ | 1.7 x 10$^{-13}$ | 9.8 x 10$^{-13}$ | 2.6 x 10$^{-13}$ | 2.7 x 10$^{-14}$ | 6.5 x 10$^{-14}$ |
| P | 3.5 x 10$^{-13}$ | 1.7 x 10$^{-13}$ | 2.4 x 10$^{-13}$ | 1.5 x 10$^{-13}$ | 5.4 x 10$^{-15}$ | 3.5 x 10$^{-14}$ |
| S | 3.6 x 10$^{-13}$ | 1.8 x 10$^{-13}$ | 2.9 x 10$^{-13}$ | 1.6 x 10$^{-13}$ | 6.8 x 10$^{-15}$ | 3.8 x 10$^{-14}$ |
| Sc | 4.4 x 10$^{-13}$ | 1.8 x 10$^{-13}$ | 5.3 x 10$^{-13}$ | 1.6 x 10$^{-13}$ | 1.3 x 10$^{-14}$ | 3.8 x 10$^{-14}$ |
| Si | 3.6 x 10$^{-13}$ | 1.8 x 10$^{-13}$ | 2.0 x 10$^{-13}$ | 1.5 x 10$^{-13}$ | 4.4 x 10$^{-15}$ | 3.3 x 10$^{-14}$ |
| Ti | 4.7 x 10$^{-13}$ | 2.0 x 10$^{-13}$ | 5.7 x 10$^{-13}$ | 2.1 x 10$^{-13}$ | 1.4 x 10$^{-14}$ | 5.1 x 10$^{-14}$ |
| V | 5.7 x 10$^{-13}$ | 2.1 x 10$^{-13}$ | 6.1 x 10$^{-13}$ | 2.2 x 10$^{-13}$ | 1.6 x 10$^{-14}$ | 5.4 x 10$^{-14}$ |
| H | 1.1 x 10$^{-14}$ | 4.7 x 10$^{-14}$ | 2.5 x 10$^{-16}$ | 1.1 x 10$^{-14}$ | 4.8 x 10$^{-18}$ | 2.4 x 10$^{-15}$ |
| He | 3.7 x 10$^{-14}$ | 7.4 x 10$^{-14}$ | 1.5 x 10$^{-15}$ | 2.2 x 10$^{-14}$ | 2.9 x 10$^{-17}$ | 4.7 x 10$^{-15}$ |
| C | 1.4 x 10$^{-13}$ | 1.2 x 10$^{-13}$ | 2.7 x 10$^{-14}$ | 6.5 x 10$^{-14}$ | 4.9 x 10$^{-16}$ | 1.4 x 10$^{-14}$ |
| O | 1.9 x 10$^{-13}$ | 1.2 x 10$^{-13}$ | 5.7 x 10$^{-14}$ | 6.5 x 10$^{-14}$ | 1.0 x 10$^{-15}$ | 1.4 x 10$^{-14}$ |
| AVERAGE | 4.1 x 10$^{-13}$ | 1.8 x 10$^{-13}$ | 4.5 x 10$^{-13}$ | 1.7 x 10$^{-13}$ | 1.1 x 10$^{-14}$ | 4.1 x 10$^{-14}$ |

## 4.3 NEW SPECIES

For the sample bombarded at 14 K it was possible to determine the formation cross-sections of the new produced species such as OCN$^-$, CO$_2$, CO, CN$^-$ and H$_2$O. For the H$_2$O this value must be employed with caution due to the contribution of residual gas. As discussed by de Barros et al. (2011) the evolution of the column density of new produced species in the samples is best described by the equation:

$$N_k = N_0 \cdot \sigma_{Fk} \cdot (F - (\frac{\sigma_d + \sigma_{dk}}{2}) \cdot F^2) \tag{8}$$

where $N_k$ is the column density of a given daughter species $k$ in molecules cm$^{-2}$, $N_0$ is the initial column density of parent species, $F$ is the fluence in ions cm$^{-2}$, $\sigma_{Fk}$ and $\sigma_{dk}$, are the formation cross-section and dissociation cross-section of the daughter species in cm$^2$, respectively, **and** $\sigma_d$ the dissociation cross-section of the parent **species**.

Figure 6 shows the chemical evolution of the sample at 14 K as a function of the fluence. We observe the decrease of glycine column density as a function of fluence in contrast with the appearing of new species produced: CN$^-$, CO, OCN$^-$, CO$_2$ and H$_2$O. For sake of comparison, the evolution of the glycine column density in the experiment performed at 300 K is also presented. The column density evolution of glycine in both experiments was described by simple relation:

$$N = N_0 \cdot \exp(-\sigma_d \cdot F) \qquad [9]$$

where $N$ is the column density, $N_0$ the initial column density in molecules cm$^{-2}$, $F$ the fluence in ions cm$^{-2}$ and e $\sigma_d$ the dissociation cross-section in cm$^2$. The red curves in **Figure 6** are the best fit obtained by employing Eq. [8] **at the data of daughter species and [9] at the glycine data set.**

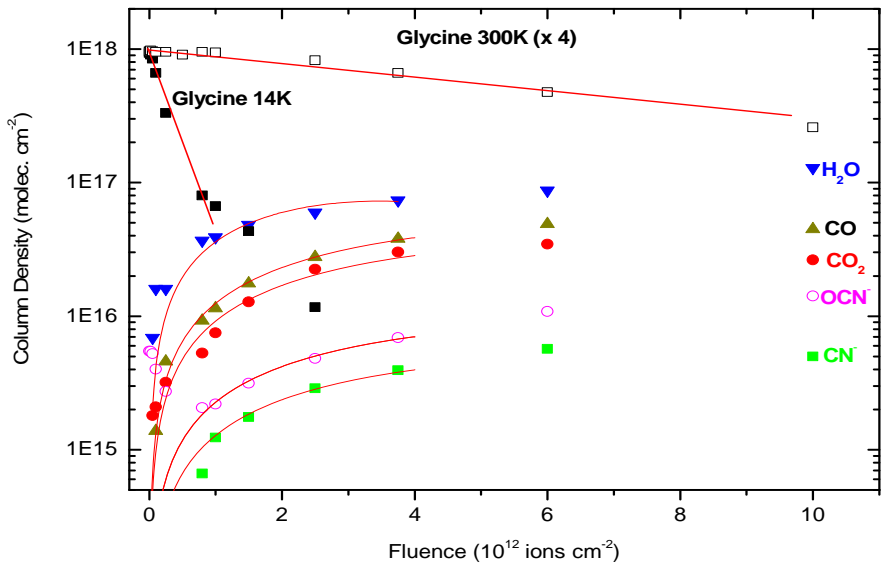

**Figure 6.** Chemical evolution of bombarded α-glycine by $^{58}$Ni$^{11}$+ ions at 14 K. For sake of comparison, the evolution of the glycine column density in the experiment performed at 300 K is also presented. The red traces are the best fit obtained by Eq. [7] and [8].

From Figure 6 we **also** observe that the water abundance, as well as other daughter species, increases continually up to the end of the experiment. It is interesting to note that at the fluence of approximately 1.5 x 10$^{12}$ ions cm$^{-2}$ the amount of produced water is very similar of the amount of remaining glycine (~5 x 10$^{16}$ cm$^2$ = 5% of glycine in the beginning of the experiment). For fluencies higher than that, the sample composition does not show significant changes (excepting the large decrease of glycine as a function of the fluence). The decrease in the glycine column density follows the Eq. [9]. However at low temperature this is only valid up to fluences around 1 x 10$^{12}$ ions cm$^{-2}$. For higher fluence, we observe that the sample is slightly less destroyed as function of fluence than in the beginning of the experiment. A similar behavior has also been observed by Pilling et al. (2013) in the experiments bombarding α-glycine at 300 K with 1 MeV protons at the fluence higher than 10$^{14}$ ions cm$^{-2}$. Such behavior may be explained by some compaction effect in the sample, changes in crystalline structures or newly produced glycine from its daughter species.

This issue could be investigated by employing isotopic labeling of glycine (e.g. $NH_2CH_2^{13}COOH$) in future experiments.

The $CN^-$, $CO_2$ and CO molecules trapped in ice reached a maximum production at about $6.0 \times 10^{12}$ ions $cm^{-2}$. For fluences higher than that, their abundance decreased. As observed with the water molecule, the new formed $OCN^-$ increased as function of the fluence. At fluence around $1.0 \times 10^{12}$ ions $cm^{-2}$, about 93% of the glycine molecules (at 14 K) were dissociated by action of the cosmic rays and became daughter species ($CN^-$, CO, $OCN^-$, $CO_2$ and $H_2O$). The other 7% may have suffered induced sputtering by impact of the heavy ions. The sputtering of bombarded frozen samples by heavy ions was extensively described previously in the literature (Seperuelo Duarte et al., 2009; Pilling et al. 2010a, b).

The measurement of the column density of water in the experiments is a difficult task because the IR bands of glycine also lay in the range of water bands. However, the following methodology was employed to estimate the column density of water: we integrated only the left side of the infrared absorbance profile of water (among the wave numbers 3628 $cm^{-1}$ and 3300 $cm^{-1}$) and multiplied it by a factor of two. The left side of this band is virtually free from bands of other species present the current experiment.

**Table 5.** Formation cross-sections of the newly produced species from the bombardment of glycine by 46 MeV $Ni^{11+}$ at 14 K. The estimated experimental error was below 10%.

| Daughter Species | $\sigma_{Fk}$ ($10^{-14}$ $cm^2$) | $\sigma_{dk}$ ($10^{-13}$ $cm^2$) |
|---|---|---|
| $CN^-$ | 0.1 | <0.1 |
| $OCN^-$ | 0.2 | <0.1 |
| $CO_2$ | 1.0 | <0.1 |
| CO | 1.3 | <0.1 |
| $H_2O$* | ~4 | ~2 |

*The cross section for $H_2O$ should be used with caution due to an eventual contamination by the residual gas.

Table 5 shows the formation and dissociation cross-sections of some new species ($CN^-$, $OCN^-$, $CO_2$, CO and $H_2O$) identified during the bombardment of the α-glycine crystal at 14 K. The values were obtained by employing Eq. [8] on experimental data. The cross section for $H_2O$ should be used with caution due to an eventual contamination by the residual gas. The production of water from the bombardment of glycine species may also be related with the formation of peptide bonds but other chemical pathways can also lead to the water formation. Future experiments with isotopic labeling using glicina with $^{18}O$ will help to clarify this issue.

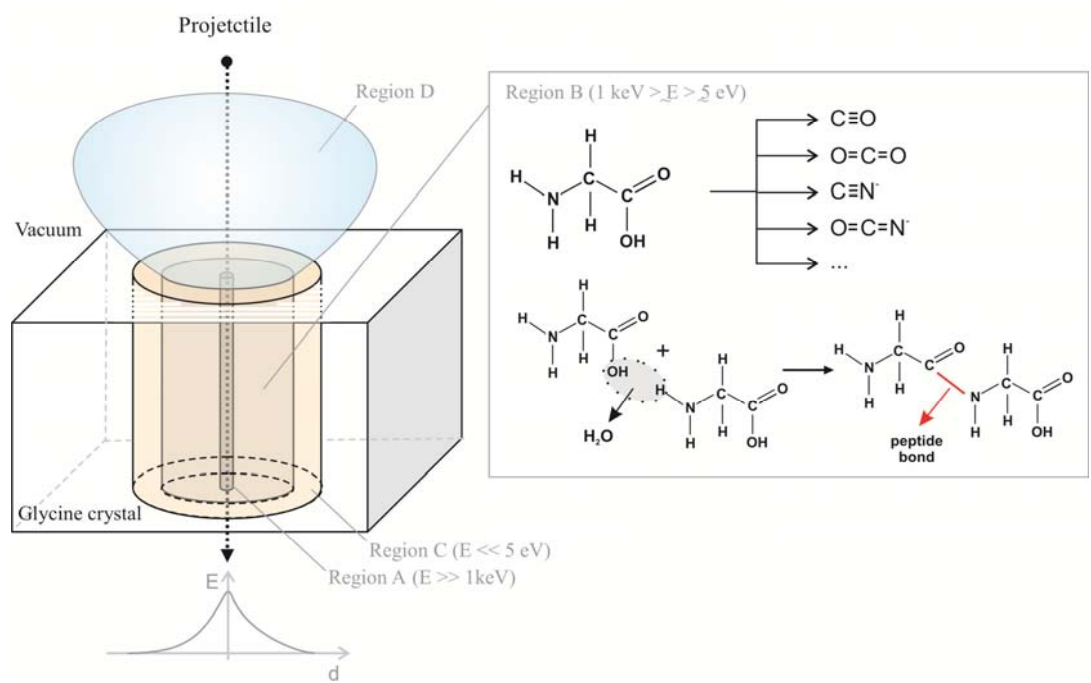

**Figure 7**. Schematic diagram of the three different physical-chemical regions surrounding the ion track during the bombardment of glycine sample (named as region A, B, C and D). Selected reaction pathways for some daughter specie that can occur in region B and a schematic plot of the energy delivered within the sample, as a function of distance of the ion track, are also shown. Tentative reaction for an eventual production of peptide bond is also illustrated. See details in the text.

As discussed by Andrade et al. (2008) and Pilling et al. (2010a), the energy delivered by the incoming projectiles (about 46 MeV ~ 7.3 x $10^{-12}$ J) induce several physical-chemical changes in the target. Basically, they produce 4 main distinct physical-chemical regions in the target due to the radial energy gradient centered at the ion track: Region A (track region) – A small cylinder-like shaped region (d ~ 3Å) along the ion track in which the energy delivered is so high that all the molecules of the target are completely atomized (for energies E >> 1 keV); Region B (ionization/dissociation region) – a region (d ~ 30 - 100 Å), surrounding the region A, in which the energy is not enough to atomize the target but induce bond ruptures (dissociation) and electron loss (ionization) (for energies 1 keV $\gtrsim$ E $\gtrsim$ 5 eV); The emission of secondary electrons inside this regions is also another source of energy input for chemical reactions. After the ion bombardment, the molecules within this region can be converted to radicals that may react to produce news species such as CO, $CO_2$, $OCN^-$ and amides. Moreover, the energy available there can be high enough to allow some reaction, that have high values of activation barrier to occur, such as the formation of peptide bonds between two amino acids; Region C (morphological changes region) - A region far from the ion track (d > 1000Å) in which the energy available is not so much, inducing only morphological changes in the sample such as changing in the crystalline structure and intermolecular bond ruptures (for energies E << 5 eV); Finally, Region D, a plume-like shaped region outside the surface which contains ejected ionic and atomic species, radicals and molecular clusters sputtered from the sample. Some reactions in gas phase may occur inside this region, depending of the number density and energy of the fragments. Details about the production of molecular clusters in this region can be obtained at Collado et al. (2004) and Andrade et al. (2008). Figure 7 shows a schematic diagram of these four different physical-chemical regions surrounding the ion track during the bombardment of glycine sample. Selected reaction pathways for some daughter species as well the suggested peptide bond formation that can occur in region B and a schematic plot of the energy delivered within the sample, as a function of distance of the ion track, are also shown. The peptide bond occurs by the link between the carboxyl group of the one amino-acid and the amine group of another amino-acid, forming an amide group with the release of a water molecule. Such intermolecular bond between amino acids is the base for the production of longer molecular chains including proteins.

Possible evidence of peptide bonds formation was also suggested by Pilling et al. (2013) in similar experiments employing light ions cosmic ray analogs (1 MeV protons) by the appearing of a peak around 1650 cm$^{-1}$ in the IR spectra of the bombarded glycine samples, which was attributed to amide bond. In addition, according to Kaiser et al. (2013), the formation of amino acids polymers, especially dipeptides, is of crucial relevance to the pre-biotic processes that preceded the onset of life on Earth. Once delivered to the planet by meteorites and comets (Oró, 1961), dipeptides could have acted as catalysts in the formation of sugars and enzymes (Weber & Pizzarello, 2006). Kaiser et al. (2013), in an experiment simulating interstellar analog ices at 10K containing mixture of different gases (e.g. carbon dioxide, ammonia and hydrocarbons) bombarded by energetic electrons, also found intense absorptions profiles in the IR spectra, among wavenumbers ~1600 cm$^{-1}$ and 1700 cm$^{-1}$ attributed to amide bonds. The chromatographic analysis of organic residues from such experiments at room temperature have shown nine different amino acids and at least two dipetides Gly-Gly and Leu-Ala.

A comparison between the infrared spectra (from 1800 cm$^{-1}$ to 750 cm$^{-1}$) of the residues produced after the bombardment of 1 x 10$^{13}$ Ni ions cm$^2$ at the two glycine samples is shown in Figure 8. The two infrared profile presented were adjusted by a sum of nine different Gaussians, that represents different infrared profiles of molecular vibration modes. The Gaussian profiles centered around 1650-1670 cm$^{-1}$ at 1590-1690 cm$^{-1}$ and at ~ 1320 cm$^{-1}$ were tentatively assigned to the amide I, amide II and amide III bands, respectively, following discussion presented by Kaiser et al. (2013) and by Reis et al. (2006).

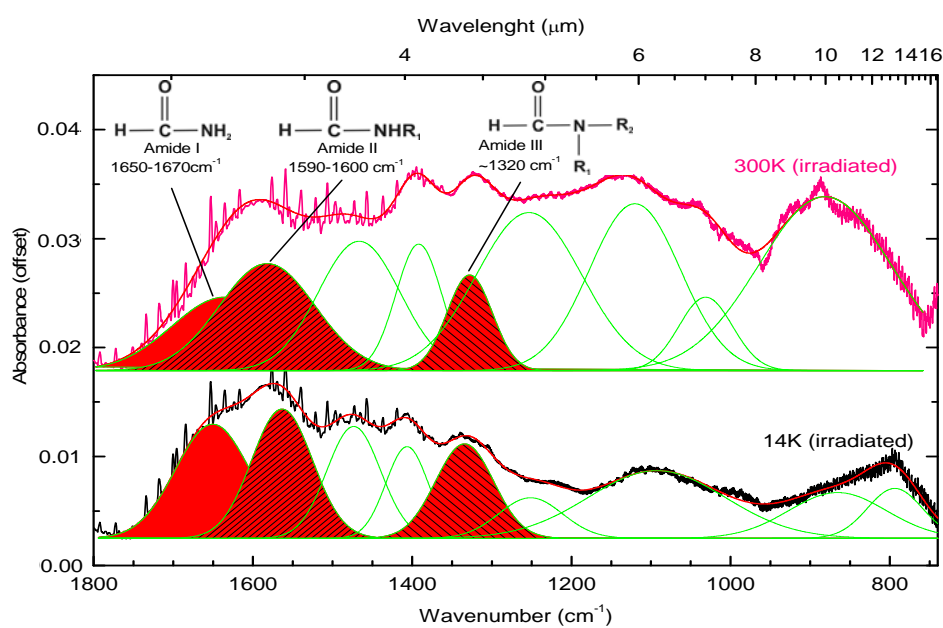

**Figure 8.** Expanded view of infrared spectra of the residues produced after the bombardment employing 1 x 10$^{13}$ Ni ions cm$^2$. Filled curves presents a tentative assignments of amide bands obtained from a spectral deconvolution employing nine Gaussian profiles, in the range from 1800 cm$^{-1}$ to 750 cm$^{-1}$.

## 4.4 HEATING OF THE SAMPLE

After the bombardment by heavy ions, the sample at 14 K was slowly heated up to room temperature while in high vacuum conditions. During this process, several infrared spectra were taken to evaluate the chemical changes during heating. Examples of these spectra are presented **in Figure 9.** This Figure shows the evolution on the IR spectra from 3550 to 600 cm$^{-1}$ of α-glicina sample, bombarded at the fluence of 1x10$^{13}$ ions cm$^{-2}$ at 14 K, during heating to room temperature. The temperature of each spectrum is indicated. For sake of comparison, the upper spectrum shows data of the non-bombarded sample at 14 K. The band associated to CO (~2137 cm$^{-1}$) is the first band to diasappear during the heating, indicating the fast evaportation of CO. This species was completely desorbed at 100 K. Some cristaline changes in the water profile at 3300 cm$^{-1}$ is also observed at the spectrum obtained at 100K. The spectrum at 180 K shows that molecules such as $CO_2$ (~2336 cm$^{-1}$) and $CN^-$ (2080 cm$^{-1}$) are also completely desorbed from the sample and water represents only a small fraction of the sample. During the sample heating, the $OCN^-$ band, initialy at 2165 cm$^{-1}$ in the spectrum at 14 K, seems to shift to shorter wavenumber (longer wevelenghts). This behavior was also observed by Pilling et al. (2010a) in the $H_2O:NH_3:CO$ (1:0.6:0.4) ice bombarded in similar conditions. The authors suggest that such a band observed at higher temperaures may be associated with a non-volatile aliphatic isocyanide R-N≡C (~2150 cm$^{-1}$).

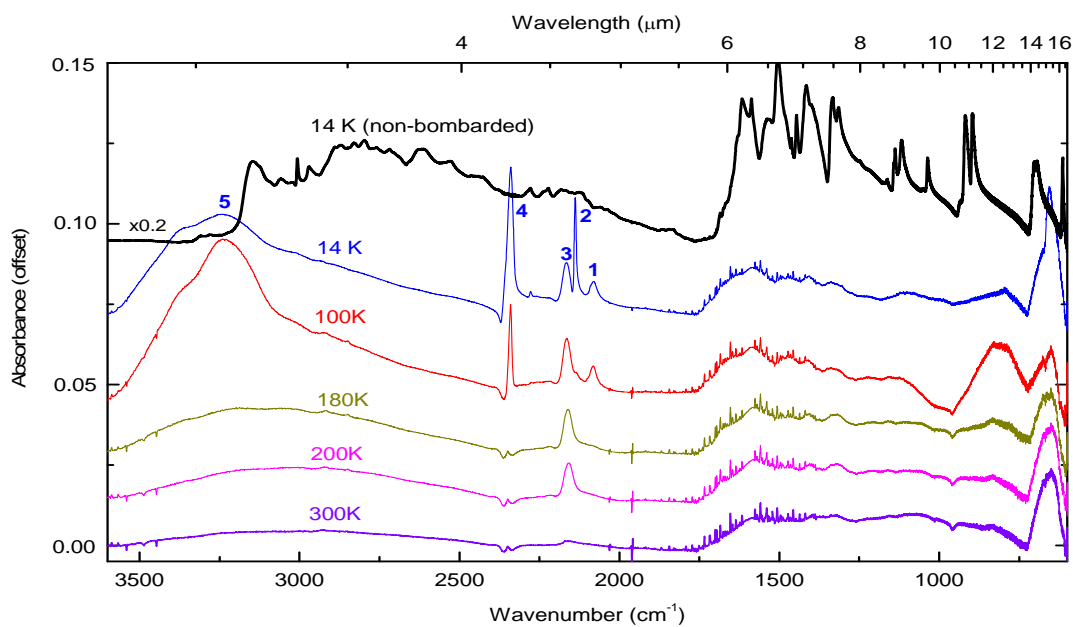

**Figure 9.** Infrared spectra of bombarded glycine at the fluence of 10$^{13}$ ions cm$^{-2}$ during heating from 14 K to 300 K. For sake of comparison the upper spectrum shows the data of the non-bombarded sample at 14K. The peak of daughter species formed on the irradiated sample at 14 K are indicated by numbers: 1 – $CN^-$ (2080 cm$^{-1}$); 2 – CO (2137 cm$^{-1}$); 3 - $OCN^-$ (2165 cm$^{-1}$); 4 - $CO_2$ (2336 cm$^{-1}$); 5 - $H_2O$ (3280 cm$^{-1}$).

Figure 10 presents the spectra of bombarded glycine at the fluence 1 x 10$^{13}$ ions cm$^{-2}$ at 14 K (second lowermost curve) and 300 K (uppermost curve). For comparison purpose, spectrum of bombarded glycine at 14 K following heating to 300 K (middle curve) and spectra of virgin glycine at 14 K (lowermost curve) and 300 K (second uppermost curve) are also given.

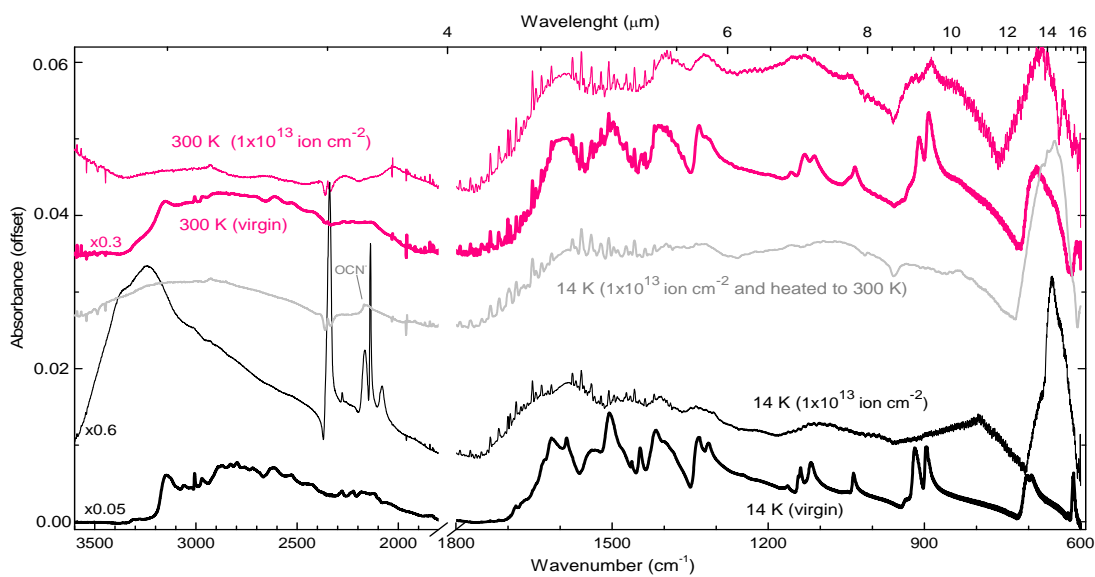

**Figure 10.** Spectra of bombarded glycine at fluence 1 x 10$^{13}$ ions cm$^{-2}$ at 14 K (bottom curve) and 300 K (upper curve). For comparison purpose, spectrum of bombarded glycine at 14 K following heating to 300 K (middle curve) and of virgin sample at 14 K and 300 K is also given. See details in text.

As discussed previously, the CO (~2137 cm$^{-1}$), CN$^-$ (2080 cm$^{-1}$), CO$_2$ (~2336 cm$^{-1}$) and H$_2$O (~3280 cm$^{-1}$), formed at 14 K, are not observed in the spectrum of processed glycine heated to 300 K (as can be seen in the Figure 9), suggesting they were desorbed. The molecule OCN$^-$ was not observed in the sample bombarded at 300 K, however, this specie is present, in small amount, on the spectrum of the bombarded sample heated to 300 K. This suggest that a fraction of OCN$^-$ (or other XCN$^-$ species), may be trapped inside the crystaline structure of the bombarded sample at 14 K and still is there after sample heating to 300 K. Such a possible trapping scenario of new produced species triggered by ion bombardment in ices was simulated theoretically by Anders & Urbassek (2012).

## 5 ASTROPHYSICAL IMPLICATIONS

The delivery of the simple organic species found in the ISM such as CH$_4$, H$_2$CO, CH$_3$OH onto Early Earth, by comets and meteorites, has been suggested as the most important source for formation of the prebiotic organic compounds such as amino acids (e.g. Cronin, 1998; Glavin & Dworkin, 2009). There is also a long-standing hypothesis in the literature that molecular precursors important for the origin of the life on Earth were first formed in space and subsequently delivered to Earth through impact events (Oró, 1961; Chyba and Sagan, 1992). Glycine, the simplest amino acid found on the proteins of all life forms of the Earth

wasn´t found yet in interstellar medium. According to Holtom et al., (2005), an upper limit for column density of this species in molecular clouds could be about $10^{12}$-$10^{14}$ molecules cm$^{-2}$, but the amount of interstellar glycine in gas phase is still unknown. It is an open question whether glycine is abundant enough to become crystaline in space environments. Bersntein et al. (2002), by means of a laboratory demonstration, showed that glycine can be produced from ultraviolet photolysis of the analogues of icy interstellar grains. The stability of glycine in the presence of ionizing space agents have beem studied by several research groups, for example involving UV (e.g. Guan et al, 2010; Peeters et al., 2003; Ehrenfreund et al., 2001; ten Kate et al., 2005; Ferreira-Rodrigues et al., 2011), X-Rays (e.g. Pilling et al., 2011b), electron beam (Abdoul-Carime & Sanche 2004) and fast ions (Gerakines et al., 2012, Pilling et al., 2013). For this, the space environments are simulated in laboratory, for the survival of molecule to be investigated during the formation of the Solar System (SS) and the Early Earth.

The flux of the galactic cosmic ray ions in the ISM can be obtained by equation

$$\Phi_z(E) = \frac{C_z E^{0.3}}{(E+E_0)^3} \quad [cm^{-2} s^{-1} (MeV/u)^{-1}] \quad [10]$$

given by Webber & Yushak (1983), based on the measurements of $_1$H, $_2$H, $_3$He and $_4$He made from balloon and *Voyager*. $E_0$ is a form parameter between 0 and 940 MeV. Changes in $E_0$ will change the spectra of low-energy cosmic rays substantially, but have almost no effect for the high-energy end. Smaller values of $E_0$ represent more low-energy cosmic-ray particles (Shen et al., 2004). Webber & Yushak (1983) found that $E_0$ = 300±100 MeV can explain the measured $^3$He/$^4$He ratio and their observed spectra very well. We assumed $E_0$ = 400 MeV, determined by Shen et al. (2004) as the best value for estimating the fluxes. In this equation, $C_z$ is the normalization constant that can be estimated for all ions as function of abundance of H on the cosmic rays of $10^6$ atoms and normalization constant $C_1$ = 9.42 x $10^4$. Employing the ion abundances taken from Drury, Meyer & Ellison (**1999**), we calculated the normalization constant for the other ions by $C_z$= 9.42 x $10^4$ ($n_z/10^6$). Here, $n_z$ indicates the number density of a given atom of atomic number Z as function of the number density of hydrogen atoms. For atoms that there are not data about the number density are unavailable (e.g. Cl, K, Sc, Ti, V, Cr and Mn) we used an average. The constant $C_z$ of each ion as function the hydrogen abundance and the respective relative abundances in the ISM are presented in Table A1 (appendix).

Inside the SS, the flux density of solar wind plus solar energetic particles of cosmic rays can be estimated by equation given by De Barros et al. (2011)

$$\Phi_z(E) = A_1 \exp(-\eta E) + \frac{A_2}{(E)^2} \quad [cm^{-2} s^{-1} (MeV/u)^{-1}] \quad [11]$$

where $A_1$ and $A_2$ are parameters used to estimate the flux density of solar energetic particles in regions of low and high energy, respectively, according the energy range given from De Barros et al. (2011). In this work, only the regions of highest energies were studied. So, the $A_1$ parameter was disregarded. The $A_2$ was estimated for all ions in function of abundance and of $A_2$ of H in the solar wind given by De Barros et al. (2011), being 3 x $10^8$ atoms and $A_2$= 46, respectively. With the abundance values of ions in the SS, from Drury, Meyer & Ellison (**1999**), we normalize these values as function of the number density of hydrogen (3 x $10^8$ atoms) and estimate $A_2$ of all other ions by $A_{2z}$= 46 ($n_z$/3 x $10^8$). Where $n_z$ indicates the number density of a given atom of atomic number Z as function of the number density of hydrogen atom in the SS. The $A_2$ values and the hydrogen relative abundance in the SS are **also** presented in Table **A1 (appendix).** The variable *E* is the energy studied on the range ~0.1-1.5 x $10^3$ MeV/u and η is the parameter about 2400 for He, Fe and O (De Barros et al.,2011).

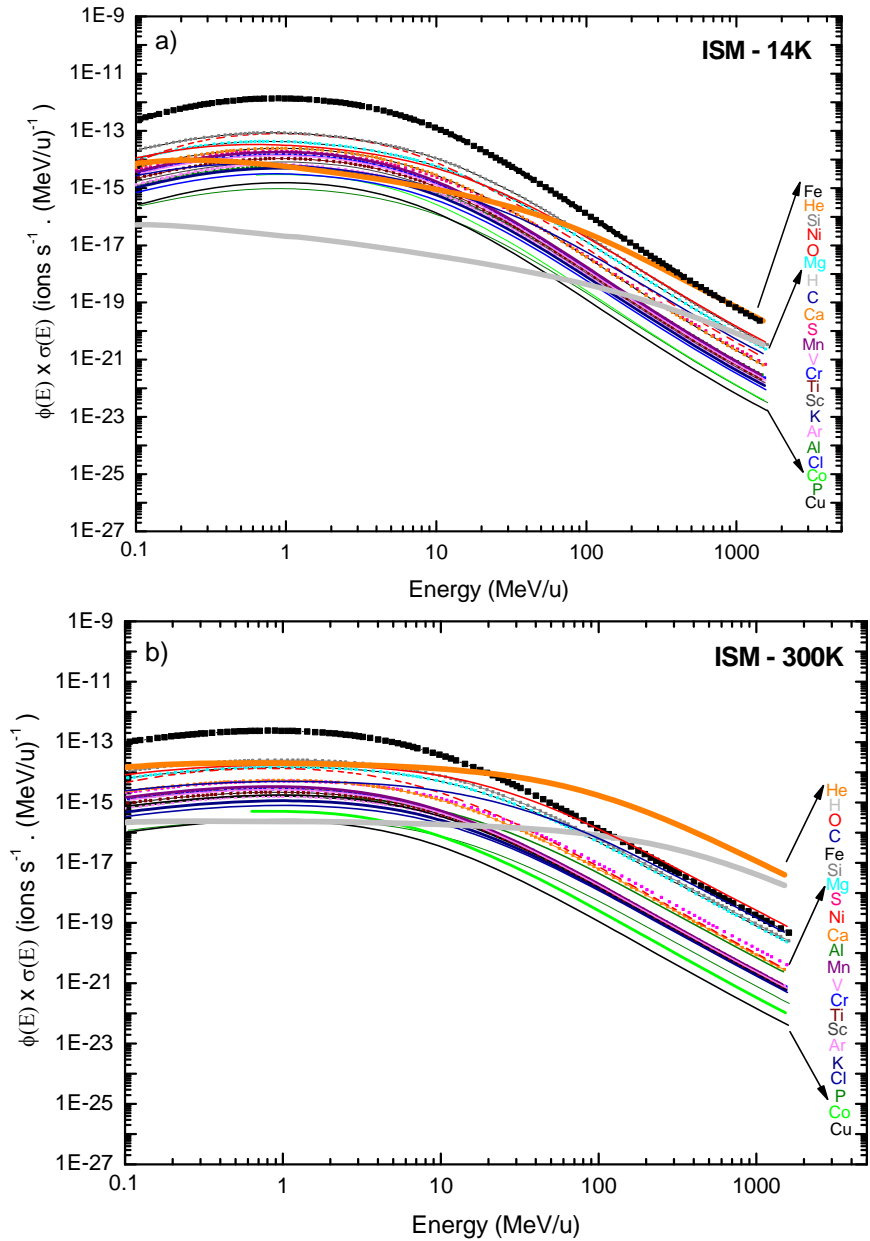

**Figure 11.** Differential dissociation rate of glycine by cosmic rays in the interstellar medium (ISM) as a function of the energy.
a) For the experiment at 14 K and b) for 300 K.

Once $\Phi_z(E)$ e $\sigma_{d,z}(E)$ known, the half-lives (τ) of the glycine molecule in the ISM and SS, at 14 and 300 K, due to cosmic rays irradiation could be evaluated using equation [12] (see Table 7). The half-life of glycine, by Andrade et al. (2013), for the energy range of ~0.1 – 1.5 x $10^3$ MeV/u, in the ISM was estimated to be 7.8 x $10^3$ years (300 K) and 2.8 x $10^3$ years (14 K). In the SS the values were 8.4 x $10^2$ years (300 K) and 3.6 x$10^3$ years (14 K). We estimated other values of half-lives by the same methodology, changing energy range and cosmic rays ions composition (Table 7).

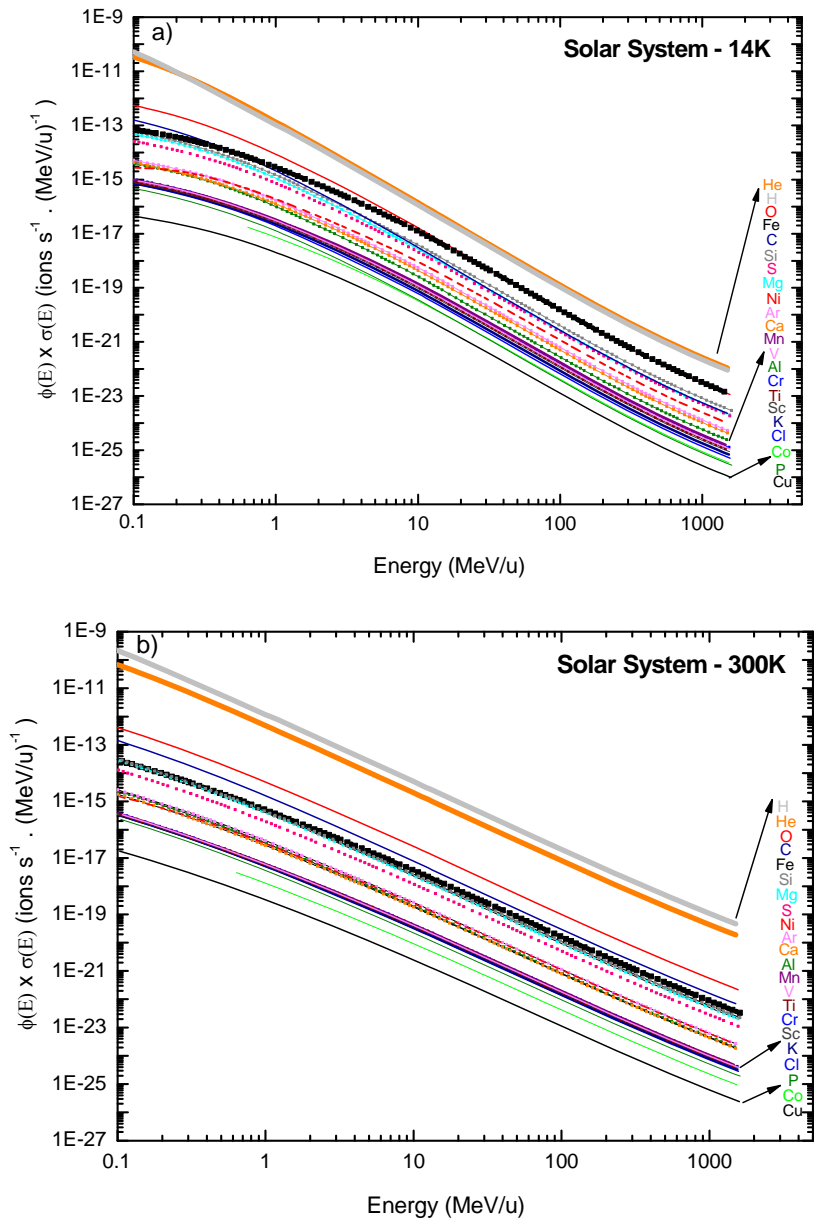

**Figure 12.** Differential dissociation rate of glycine by cosmic rays in the solar system (SS) as function of the energy. a) For the experiment at 14 K and b) for 300 K.

Figure 11a and 11b show the differential dissociation rate of glycine by cosmic rays in the ISM at 14 and 300 K, respectively. The differential dissociation rate is a parameter defined by product between the flux of ions and their respective cross sections in function of energy ($\Phi(E) \times \sigma_d(E)$). Figure 12a and 12b show the same parameter, but for the SS. According these Figures, the differential dissociation rate of glycine induced by of cosmic rays ($\Phi(E) \times \sigma_d(E)$) in the ISM predominates between approximately 0.1-10MeV/u, it is higher for the sample at 14 K that at 300 K. As already discussed, this can be due to the high density in the sample at low temperature, enabling the dissociation of higher number of molecules under impact of cosmic rays. Furthermore, dissociation of molecules by reaction among species formed with other already presents in the sample can occur. Fe (black line) is the element that predominates in the glycine dissociation, in the ISM. According to the Figures 12a and 12b, H (gray line) and He (orange line) are the ions that predominate in the molecule dissociation, in the SS. The values presented in the Table 6 show this relation between the cosmic ray composition and the energy ranges studied. The highest dissociation rate is found in the SS, because the fluxes of H and He integrated over of energy, in this environment, are higher than in the ISM. Furthermore, the glycine dissociation cross sections in relation to these ions are higher in the hottest environments (see Table 4). In addition, the work of Andrade et al. (2013) with formic acid at 15 K, irradiated by cosmic ray analogs, showed that protons and galactic He interact with deeper layers of ice. The SRIM calculations show that H and He ions have the highest perpendicular penetration depth during the bombardment. The $S_e$ of other ions increases initially, and them decrease. This can explain a higher molecule dissociation rate under action of cosmic rays including the action of H and He ions, in both temperatures (14 and 300 K) and the two energy ranges studied (~0.1-1.5 x $10^3$ MeV/u **and** ~0.1-10 MeV/u). The glycine dissociation rate decreases when these ions are not take into account, therefore increasing the half live of the molecule during the bombardment.

According to Andrade et al. (2013) with the integration of the differential dissociation rate over energy ($\int \Phi_z(E)\sigma_{d,z}(E)dE$), it is possible to estimate the half-lives of glycine in the ISM and SS, at two temperatures (14 K and 300 K), by equation

$$\tau = \frac{\ln(2)}{k} \quad [s] \tag{12}$$

where $k$ is the sum $\Sigma j \int \Phi_z(E)\sigma_{d,z}(E)dE$. $\Phi_z$ is the estimated flux density of cosmic rays *($\Phi_{HCR}(E)$)* between $E$ and $E + dE$ in ions cm$^{-2}$ s$^{-1}$ (MeV/u)$^{-1}$, and $\sigma_{d,z}$ the destruction cross section in cm$^2$, which is a function of the ion energy of a given ion **with atomic number** *Z*. The integration was performed over the whole energy range studied (~0.01 - 1.5 x $10^3$ MeV/u), over the area below of a given curve in Figures 11a, 11b, 12a and 12b. The sum $\Sigma j \int \Phi_z(E)\sigma_{d,z}(E)dE$ (k) was performed by the methodology suggested by Andrade et al. (2013) for all galactic cosmic ray ions studied including H, He, C, O and the heavy ions (12 ≤ Z ≤ 29). In this work we defined *k'* as the estimated value for the ionization rates that take into an account only the ions with energies between 0.1-10 MeV. **Table 6 presents these defined ionization rates *k* and *k'* of glycine molecule on the ISM and SS, at 14 and 300 K, exposed to cosmic rays, in the energy ranges ~0.1-1.5 MeV/u and ~0.1-10 MeV/u. For comparison purpose the ionization rates that consider only heavy ions (named as k$_{HCR}$ and k$_{HCR}$´) are also listed in this table.**

**Table 6.** Ionization rates k (in s$^{-1}$) of glycine molecule on the ISM and SS at 14 and 300 K exposed to cosmic rays, in the energy ranges ~0.1 - 1.5 MeV/u and ~0.1 – 10 MeV/u.

|  | ISM (dense clouds) | | SS | |
|---|---|---|---|---|
| **Ionization rate** | 14K | 300K | 14K | 300K |
| k | 7.92 x 10$^{-12}$ | 2.80 x 10$^{-12}$ | 6.15 x 10$^{-12}$ | 2.61 x 10$^{-11}$ |
| k' | 6.61 x 10$^{-12}$ | 1.86 x 10$^{-12}$ | 6.15 x 10$^{-12}$ | 2.61 x 10$^{-11}$ |
| k$_{HCR}$ | 7.64 x 10$^{-12}$ | 1.69 x 10$^{-12}$ | 4.77 x 10$^{-14}$ | 1.43 x 10$^{-14}$ |
| k$_{HCR}$' | 6.46 x 10$^{-12}$ | 1.51 x 10$^{-12}$ | 4.77 x 10$^{-14}$ | 1.43 x 10$^{-14}$ |

k: Ionization rate considering the most abundant ions (12 ≤ Z ≤ 29, H, He, C and O) with energy between ~0.1-1.5x10$^3$ MeV/u. k': Ionization rate considering the most abundant ions (12 ≤ Z ≤ 29, H, He, C and O) with energy between ~0.1−10 MeV/u. k$_{HCR}$: Ionization rate considering only heavy ions (12 ≤ Z ≤ 29) with energy between ~0.1-1.5x10$^3$ MeV/u. K$_{HCR}$': Ionization rate considering only heavy ions (12 ≤ Z ≤ 29) with energy between ~0.1−10 MeV/u.

**To make easier the comparison between the current data with previous works,** we also calculated, the half-life ($\tau^*$) of α-glycine crystals, exposed to heavy ions in the energy range of ~0.1-10 MeV, employing the equation [13] by Pilling et al. (2010a, b),

$$\tau^* = \frac{\ln(2)}{\phi \cdot \sigma_d} \qquad [13]$$

where $\phi$ is the flux of heavy ions cm$^{-2}$ s$^{-1}$, estimated by Pilling et al. (2010a, b) and $\sigma_d$ is the dissociation cross-section in cm$^2$ determined in this work for 14 K and 300 K. As discussed by Pilling et al. (2010a, b), an estimative for the flux of heavy ions (12 ≤ Z ≤ 29) with energy between 0.1−10 MeV/u in the interstellar medium is about 5 x 10$^{-2}$ cm$^{-2}$ s$^{-1}$. Inside the SS, at Earth orbit, the flux of such heavy ions (including heavy component of cosmic rays plus energetic solar particles) in the same energy range is about 2 x 10$^{-2}$ cm$^{-2}$ s$^{-1}$. In this methodology, as consequence of higher dissociation cross sections of the coldest sample, the half-life of glycine at 300K was higher than of the glycine at 14 K. A variation of such stability with temperature was also verified by Gerakines et al. (2012). The results obtained in this work, involving the glycine cross section for Ni ions with 46 MeV only and the results obtained by Pilling et al. (2010a,b), suggest that hottest regions in ISM or SS could be a better target for glycine search in the space environments than the coldest regions. However, the half lives estimated by Andrade et al. (2013) involve other variables, such as the flux and cross section of each ion constituent of cosmic rays as function of energy, showing that the glycine survive more time in the hottest environment of ISM, under action of all ions. In the SS, the molecule can be better found in the coldest regions, under bombardment of all ions, and in hottest regions under action of heavy ions only, without the action of H and He ions.

Table 7 lists the values of half-lives $\tau$ (in years) of glycine molecule on the ISM and SS, at 14 and 300 K exposed to cosmic rays irradiation. For comparison purpose half-lives determined by Pilling et al. (2010a,b), named as $\tau^*$, are also given. At last, a correction factor ($\tau/\tau^*$) was calculated to correct the half-lives estimated by Pilling et al. (2010a, b) methodology.

**Table 7.** Half-lives τ (in years) of glycine molecule on the ISM and SS, at 14 and 300 K exposed to cosmic rays irradiation. For comparison purpose half-lives determined by Pilling et al. (2010a,b), named as $\tau^*$, are also given. At last, a correction factor ($\tau/\tau^*$) was calculated to correct the half lives estimated by Pilling et al. (2010a, b) methodology.

|  | ISM (dense clouds) | | SS | |
|---|---|---|---|---|
| **Half-life** | 14K | 300K | 14K | 300K |
| $\tau$ | $2.8 \times 10^3$ | $7.8 \times 10^3$ | $3.6 \times 10^3$ | $8.4 \times 10^2$ |
| $\tau'$ | $3.3 \times 10^3$ | $1.2 \times 10^4$ | $3.6 \times 10^3$ | $8.4 \times 10^2$ |
| $\tau_{HCR}$ | $2.9 \times 10^3$ | $1.3 \times 10^4$ | $4.6 \times 10^5$ | $1.5 \times 10^6$ |
| $\tau_{HCR}'$ | $3.4 \times 10^3$ | $1.4 \times 10^4$ | $4.6 \times 10^5$ | $1.5 \times 10^6$ |
| $\tau/\tau_{HCR}$ | ~1.0 | 0.6 | $7.8 \times 10^{-3}$ | $5.6 \times 10^{-4}$ |
| $\tau^*$ | $1.8 \times 10^5$ | $1.3 \times 10^6$ | $4.6 \times 10^5$ | $3.2 \times 10^6$ |
| $\tau_{HCR}'/\tau^*$ | 0.02 | 0.01 | 1.0 | 0.5 |
| $\tau/\tau^*$ | 0.02 | 0.006 | 0.008 | 0.00026 |

$\tau$: Half-life taking into an account the most abundant ions ($12 \leq Z \leq 29$, H, He, C and O) with energy between ~0,1-1.5x10³ MeV/u. $\tau'$: Half-life considering the most abundant ions ($12 \leq Z \leq 29$, H, He, C and O) with energy between ~0.1−10 MeV/u. $\tau_{HCR}$: Half-life by interaction heavy ions ($12 \leq Z \leq 29$) with energy between ~0.1-1.5x10³ MeV/u. $\tau_{HCR}'$: Half-life by interaction heavy ions ($12 \leq Z \leq 29$) with energy between ~0.1−10 MeV/u. $\tau^*$: half-life determined employing previous methodology (pilling et al 2010a; 2010 b).

**Table 8 shows a compilation with t**he half-lives values of α-glycine molecule in the ISM (dense clouds) under the **action of a single ionizing agent** obtained by different authors **(Ferreira-Rodrigues et al. 2011; Pilling et al. 2011a; Gerakines et al. 2012). From this table** we can conclude that if we consider the glycine dissociation cross section under action at one ionizing agent only, the **half-lives** increase. Exceptions happen for the α-glycine under bombardment of soft x-rays, due to the high dissociation cross section of molecule in this energy range (Pilling et al., 2011b).

**Table 8.** Half- lives in years (τ) of glycine molecule in the ISM (dense clouds) under action of ionizing agents only, estimated by others authors, for comparison with the half lives obtained in this work.

| Ionizing Agent | Sample | τ (dense clouds) | Ref. |
|---|---|---|---|
| UV Lamp (10 eV) | α-glycine negatively charged (300K) | $2.4 \times 10^6$ | [1] |
| Soft X-Rays (150 eV) | α-glycine (300K) | $1.8 \times 10^2$ | [2] |
| Proton (0.8 MeV) | α-glycine (300K) | $7.6 \times 10^6$ | [3] |
| $^{58}Ni^{11+}$ (46 MeV) | α-glycine (14K) | $>3.0 \times 10^6$ | [4] |
|  | α-glycine (300K) | $>1.5 \times 10^6$ | [4] |

[1] Ferreira-Rodrigues et al. (2011); [2] Pilling et al. (2011a); [3] Gerakines et al. (2012); [4] This work.

## 6 CONCLUSIONS

In this work we presented an experimental study about the stability of the amino acid glycine in a solid phase (crystalline α-glycine form) bombarded by 46 MeV $^{58}Ni^{11+}$ ions (~0.8 MeV/u). The samples at 14 and 300 K were exposed up to the fluence of $1 \times 10^{13}$ ions cm$^{-2}$ at the GANIL heavy ion accelerator, in an attempt to simulate the effects of the heavy and energetic cosmic rays in astrophysical environments. The samples were analyzed *in-situ* by Fourier Transformer infrared spectrometer. The dissociation cross-sections of glycine molecules were determined and half-lives for this species (hypothetically present) in different space regions and energy ranges were estimated. Our main results and conclusions are the following:

i) The dissociation cross section values were obtained for 14 K ($\sigma_d = 2.4 \times 10^{-12}$) and 300 K ($3.4 \times 10^{-13}$ cm$^2$). The estimated error was about 20% and 50% for the experiments at 14 K and 300 K, respectively. Considering the average value between the different specific bond rupture cross sections showed that at 14 K the glycine dissociation cross-section was about seven times higher than at room temperature. It was possible to see also, that for any adopted vibration mode or considering the average values, the glycine dissociation cross sections are higher for the sample at 14 K.

ii) The formation and dissociation cross-sections of new species formed by bombardment of the sample at 14 K were also obtained. They are respectively: CN$^-$ (0.14 x 10$^{-14}$ cm$^2$; <0.1 x 10$^{-13}$ cm$^2$), CO (1.30 x 10$^{-14}$ cm$^2$; <0.1 x 10$^{-13}$ cm$^2$), OCN$^-$ (0.24 x 10$^{-14}$ cm$^2$; <0.1 x 10$^{-13}$ cm$^2$), CO$_2$ (0.98 x 10$^{-14}$ cm$^2$; <0.1 x 10$^{-13}$ cm$^2$) and H$_2$O (4.20 x 10$^{-14}$ cm$^2$; <1.6 x 10$^{-13}$ cm$^2$). In the sample at 300 K this species could have been formed, but, they are not observed in the spectrum, suggesting they were desorbed. In the spectrum of glycine heated to 300 K this species also are not observed.

iii) The half-lives of glycine extrapolated to ISM were estimated to be 7.8 x 10$^3$ years (for 300 K sample) and 2.8 x 10$^3$ years (14 K). In Solar System the values were 8.4 x 10$^2$ years (300 K) and 3.6 x10$^3$ years (14 K). This indicates that the sample at higher temperature (in ISM) have a higher half-life. The same relation was showed by Pilling et al. (2014) for glycine molecules in the presence of fast electrons. In the hottest environments of SS, including the most abundant ion constituents of cosmic rays (heavy ions (12≤Z≤29), H, He, C and O ions), the glycine half life is lower. This happens because the effect of heavy ions in the ISM is dominant for cosmic rays. Furthermore, the dissociation rate is higher at 14 K than 300 K, in the ISM. In the SS the effects of light ions, specifically of H and He, coming mainly from the Sun, become more important. The effects of these ions are more intense inside the SS at Earth orbit and the glycine dissociation cross sections for H and He ions are highest in hottest environments of the SS. In view of these results, we can conclude that the glycine would survive during the formation of the SS under constant bombardment of cosmic rays only if the molecule was protected for example, inside of cometary surfaces or of interstellar grains. Furthermore, if the sample had thickness higher than the maximum ion penetration depth (e.g. d > 17.5 μm in considering only 46 MeV Ni ions), is possible that the deepest layers of an astrophysical ice with glycine were not bombarded by ionizing particles. So, the glycine would survive longer that the time estimation of the half-life of molecules in direct contact with the radiation field. According Pilling et al. (2014) the question that remains is: Does glycine form on grains in interstellar regions and persist to solar system formation, or is it a molecule formed and destroyed many times during the star and planet formation cycle? This work suggest also, that due to the low resistance of the glycine sample inside radiation field of SS, a considerable fraction could have been produced later in the SS evolution and persisted to seed the origins of life.

iv) Comparing two distinct methodologies about the contribution of different cosmic ray constituents in the destruction of glycine in space analog environments (e.g Andrade et al. (2013) and Pilling et al. (2010a; 2010b), it is possible to correct the values of half lives by factors ($\tau/\tau^*$) (Table 7). These factors will can be used in future investigations to correct half lives of glycine estimated by Pilling et al. (2010a, 2010b) methodology.

v) The production of water in the sample during the ion bombardment may be related with the formation of peptide bond. In addition, the possible identification of amide vibrational modes around 1650 cm$^{-1}$ and 1590 cm$^{-1}$ in the infrared spectra of the bombarded samples reinforce this scenario of peptide bond formation. However, since several other chemical pathways may lead to the formation of water (including residual gas deposition in the substrate) and, also to the formation of molecules containing amide groups, the peptide bond formation suggested here must be employed with caution. Future experiments, for example, using isotopic labeling may help top clarify this issue.

It is believed that α-glycine would be present in space environments that suffered aqueous changes such as the interior of comets, meteorites and planetesimals as discussed by Pilling et al. (2013). Therefore, it could contribute to the start of pre biotic chemistry on our planet, if glycine molecules were protected against constant bombardment by cosmic rays for example inside cometary surfaces or into interstellar grains. So, the study of the stability of glycine in these environments, under ions bombardment, provides further understanding about the role of this species in the pre-biotic chemistry on Earth.

**ACKNOWLEDGMENTS**

The authors acknowledge the Brazilian agencies CAPES, CNPq for financial support. S. Pilling thanks T. Madi, A. Domaracka, C. Grygiel, I. Monnet, the GANIL laboratory/CIMAP for parcial finnantial support during the measurments and Th. Been and J. M. Ramillon for technical support. We also thank Ms Alene Alder Rangel for the English revision of this manuscript.

**APPENDIX A: PARAMETERS EMPLOYED IN THE COSMIC RAY FLUX.**

**Table A1 lists some parameters employed in the model for the estimative of galactic and solar system cosmic ray.**

Table A1 - Relative abundance in the ISM and SS, normalization constant ($C_z$) and $A_2$ of several constituents of galactic cosmic rays related to abundance of H.

| Z | Element | Relative Abundance of $10^6$ H atoms (ISM) | Relative Abundance of $3 \times 10^8$ H atoms (SS) | $C_z$ (MI) | $A_2$ (SS) |
|---|---|---|---|---|---|
| 1 | H | $10^6$ | $3.00 \times 10^8$ | $9.42 \times 10^4$ | 46 |
| 2 | He | $10^{4.8}$ | $6.00 \times 10^7$ | $6.30 \times 10^4$ | 9.2 |
| 4 | C | $10^{3.4}$ | $7.54 \times 10^4$ | 236.6 | $1.16 \times 10^{-2}$ |
| 6 | O | $10^{3.5}$ | $1.90 \times 10^5$ | 297.9 | $2.91 \times 10^{-2}$ |
| 12 | Mg | $10^{2.8}$ | $1.19 \times 10^4$ | 59.4 | $1.82 \times 10^{-3}$ |
| 13 | Al | $10^{1.7}$ | $9.50 \times 10^2$ | 4.7 | $1.46 \times 10^{-4}$ |
| 14 | Si | $10^{2.8}$ | $1.19 \times 10^4$ | 59.4 | $1.82 \times 10^{-3}$ |
| 15 | P | $10^{0.7}$ | $9.50 \times 10$ | 0.4 | $1.45 \times 10^{-5}$ |
| 16 | S | $10^{1.9}$ | $4.75 \times 10^3$ | 7.5 | $7.30 \times 10^{-4}$ |
| 17 | Cl | $10^{1.0}$ | $1.19 \times 10^2$ | 0.9 | $1.82 \times 10^{-5}$ |
| 18 | Ar | $10^{1.1}$ | $9.48 \times 10^2$ | 1.2 | $1.45 \times 10^{-4}$ |
| 19 | K | $10^{1.0}$ | $1.19 \times 10^2$ | 0.9 | $1.82 \times 10^{-5}$ |
| 20 | Ca | $10^{1.6}$ | $6.00 \times 10^2$ | 3.8 | $9.20 \times 10^{-5}$ |
| 21 | Sc | $10^{1.0}$ | $1.19 \times 10^2$ | 0.9 | $1.82 \times 10^{-5}$ |
| 22 | Ti | $10^{1.0}$ | $1.19 \times 10^2$ | 0.9 | $1.82 \times 10^{-5}$ |
| 23 | V | $10^{1.0}$ | $1.19 \times 10^2$ | 0.9 | $1.82 \times 10^{-5}$ |
| 24 | Cr | $10^{1.0}$ | $1.19 \times 10^2$ | 0.9 | $1.82 \times 10^{-5}$ |
| 25 | Mn | $10^{1.0}$ | $1.19 \times 10^2$ | 0.9 | $1.82 \times 10^{-5}$ |
| 26 | Fe | $10^{2.8}$ | $9.48 \times 10^3$ | 59.4 | $1.45 \times 10^{-3}$ |
| 27 | Co | $10^{0.08}$ | $2.38 \times 10^1$ | 0.11 | $3.65 \times 10^{-6}$ |
| 28 | Ni | $10^{1.5}$ | $5.98 \times 10^2$ | 3.0 | $9.16 \times 10^{-5}$ |
| 29 | Cu | $10^{-0.33}$ | 5.98 | 0.044 | $9.17 \times 10^{-7}$ |